\documentclass[aps,twocolumn,floatfix]{revtex4}
\usepackage{epsfig}
\usepackage{graphicx}
\usepackage{dcolumn}
\usepackage{natbib}
\usepackage{amssymb}

\def\dj{\hbox{d\kern-0,347em \vrule width0,3em height1,252ex
depth-1,21ex \kern0,051em}}

\begin{document}

\title{Role of the sample thickness in planar crack propagation}

\author{Pallab Barai}
\affiliation{Department of Mechanical Engineering
Texas A\& M University
College Station, TX, USA 77843}
\author{Phani K. V. V. Nukala}
\affiliation{Computer Science and Mathematics Division,
Oak Ridge National Laboratory, Oak Ridge, TN 37831-6359, USA}
\author{Mikko J. Alava}
\affiliation{COMP Center of Excellence, Department of Applied Physics, Aalto University,
P.O. Box 14100, FIN-00076, Aalto, Espoo, Finland}
\author{Stefano Zapperi}
\affiliation{CNR - Consiglio Nazionale delle Ricerche, IENI, Via R. Cozzi 53, 20125,
Milano, Italy}
\affiliation{ISI Foundation, Via Alassio 11/c 10126 Torino, Italy}
\begin{abstract}
We study the effect of the sample thickness in planar crack front propagation
in a disordered elastic medium using the random fuse model.  We employ
different loading conditions and we test their stability with respect to crack growth.
We show that the thickness induces characteristic lengths
in the stress enhancement factor in front of the crack and in the stress transfer
function parallel to the crack. This is reflected by a thickness-dependent
crossover scale in the crack front morphology that goes from from multi-scaling to self-affine
with exponents in agreement with line depinning models and experiments.
Finally, we compute the distribution of crack avalanches which is shown to depend
on the thickness and the loading mode.
\end{abstract}
\maketitle

\section{Introduction}

Understanding the failure of disordered media still represent an open
problem for basic science and engineering. From the point of
view of statistical mechanics, the problems represents an interesting
example of the interplay between disorder and long-range interactions
and has thus attracted a wide interest in the past years \cite{alava06}.
Fracture typically displays intriguing size-effects and it
is not easy to formulate a general law that can predict the
the dependence of the failure strength on the relevant lengthscales
of the problem, such as the size of the notch, of the fracture process
zone and of the sample width and thickness \cite{vandenborn91,bazant99,sutherland99}

Scale and size dependence also permeate the analysis of crack morphologies,
that have been originally characterized by self-affine
scaling \cite{mandelbrot84,bouchaud97}, but a complete understanding
of the universality classes of the roughens exponent is still
an debated issue (for a review see \cite{bonamy11}).  The prevalent interpretation
of experimental measurements for  out-of-plane three dimensional cracks suggests
a roughness exponent value $\zeta\simeq 0.8$ for rupture processes occurring
inside the fracture process zone (FPZ), where elastic interactions
would be screened, crossing over at large scales to $\zeta \simeq 0.4$
when elastic interaction would dominate \cite{ponson06,bonamy06}.
Furthermore, the fracture surface has been shown to be anisotropic with
different scaling exponents in the directions parallel or perpendicular
to the crack \cite{ponson06}.

A particularly interesting and conceptually simpler example of crack roughening
is represented by the propagation of a planar crack. This case appears to be the ideal candidate
to test the theory that envisages the crack as a line moving through a disordered
medium \cite{bouchaud93,daguier97}. For planar cracks, the problem can be
mapped into a model for interface depinning with long-range forces \cite{schmittbuhl95,ramanathan97,ramanathan98,bonamy08,laurson10,laurson10b},
implying a self-affine front with a roughness exponent close to $\zeta=1/3$ \cite{ertas94,rosso02}
and avalanche propagation of the front between pinned configurations with scaling
exponents predicted by the theory
\cite{ramanathan98,bonamy08,laurson10}. Such results are also of importance for applications, like the failure of the interface between a substrate and a coating, or an adhesive layer, and the propagation of indentation cracks \cite{vellinga06}

Despite the theoretical understanding
is clear, interpreting the experimental results as proven to be a challenging task
\cite{schmittbuhl97,delaplace99,maloy01,maloy03,maloy06,salminen06,santucci10}.
Initial results indicated a roughness exponent in the range of $\zeta=0.5-0.6$ \cite{schmittbuhl97,delaplace99} that was definitely at odd with theoretical predictions.
Only recently it was shown that the early measurements focused on short lengthscales
where the crack front is not self-affine but obeys instead multiscaling, while
the predicted universal roughness exponent was recovered on larger lengthscales
\cite{santucci10}. Similarly, early measurements of the avalanche distribution
did not agree with theoretical predictions based on elastic line models \cite{maloy06,bonamy08}.
It was later realized that due to long-range interactions along the crack front avalanches
are decomposed in disconnected clusters whose scaling may differ from that
of the avalanches \cite{laurson10}.

Numerical simulations of discrete lattice models for fracture have been used in the past to
investigate the  short-scale disorder-dominated regime of planar crack front propagation
focusing either on quasi-two dimensional small thickness samples \cite{zapperi00,zaiser09} or on large-thickness bulk three dimensional samples  \cite{schmittbuhl03,stormo2012}. He we perform three
dimensional simulations of the random fuse model \cite{dearcangelis85} to better understand the role of thickness in a regime intermediate between two and three dimensions. We find that the thickness introduces
a characteristic length-scale that cuts off the long range interactions along the crack front.
As a consequence of this we find a thickness dependent roughening behavior with multiscaling
observed at low thickness and  self-affinity at large thickness. Furthermore, the sample thickness influences
the stability of crack propagation that also depends on the way loading is applied.  This is reflected also in
the avalanche behavior that in the stable propagation regime follows the predictions of the interface
depinning model \cite{bonamy08,laurson10}. Beside the theoretical implications,
understanding the role of thickness in planar crack could be interesting in view of applications for the delamination of coatings \cite{vellinga06}.

\begin{figure}[th]
\includegraphics[width=1.\linewidth]{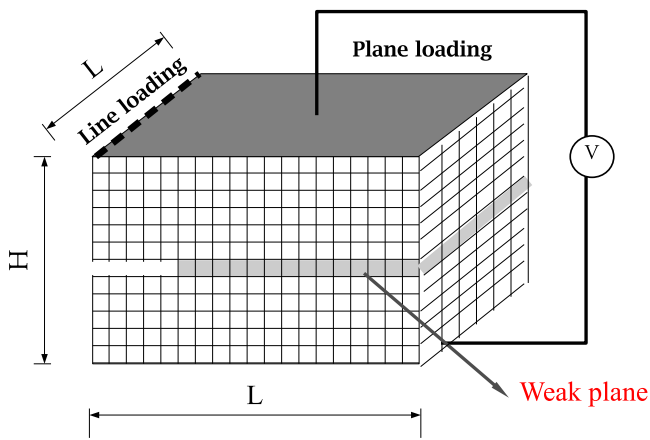}
\caption{The geometry of the model. We consider a cubic lattice of conducting bonds
with a weak plane in the middle where we place fuses with random breaking threshold.
A notch of lenghth $a$ is placed in the weak plane at the beginning of the simulation.
The voltage drop is either applied between
the top and bottom planes (plane loading) or
at the left edges of the plates, on the dotted lines (line loading).}
\label{fig:geometry}
\end{figure}

\section{The model}

Here we consider crack propagation under antiplane deformation, a scalar
problem that can be mapped to an electrical analog: the random fuse
model (RFM). In the RFM \cite{dearcangelis85} a set of conducting bonds,
with unit conductivity $\sigma_0=1$,
are arranged on a cubic lattice of size $L\times L \times H$.
To simulate the presence of a weak plane, the vertical bonds crossing the central
horizontal plane are replaced by fuses. When the local current $i_j$ overcomes a
randomly chosen threshold $t_j$, the fuse burns irreversibly.  The
thresholds are randomly distributed based on a thresholds probability
distribution, $p(t)$. In addition, an edge notch is placed on
one side of the weak plane. Periodic boundary conditions are imposed in
the direction parallel to the notch to simulate an infinite system and a
constant voltage difference, $V$, is applied between the top and the
bottom plates of the lattice (plane loading) or between two edges
(line loading). See Fig.~\ref{fig:geometry} for an illustration of the geometry.
The second boundary condition resembles the loading applied in
experiments, although here we consider mode III (antiplane shear) while
the experiments where performed under mode I (tension).

Numerically, we set a unit voltage  difference, $V = 1$, and solve the Kirchhoff
equations  to determine the current flowing in each of the
fuses. Subsequently, for each fuse $j$, the ratio between the current
$i_j$ and the breaking threshold $t_j$ is evaluated, and the bond
$j_c$ having the largest value, $\mbox{max}_j \frac{i_j}{t_j}$, is
irreversibly removed (burnt).  The current is redistributed
instantaneously after a fuse is burnt implying that the current
relaxation in the lattice system is much faster than the breaking of a
fuse.  Each time a fuse is burnt, it is necessary to re-calculate the
current redistribution in the lattice to determine the subsequent
breaking of a bond.  The process of breaking of a bond, one at a time,
is repeated until the lattice system falls apart. In this work, we
assume that the bond breaking thresholds are distributed based on a
uniform probability distribution, which is constant between 0 and 1.
An alternative would be to study power law distributions
exponents and control the disorder strength varying the
exponent, as done in Ref.~\cite{batrouni98}. Since the robustness of
the model behavior with respect of disorder has been extensively studied
in the literature we concentrate our effort on a single type of disorder,
extending the statistical sampling and the range of lattice sizes.

Numerical simulation of fracture using large fuse networks is often
hampered due to the high computational cost associated with solving a
new large set of linear equations every time a new lattice bond is
broken. Although the sparse direct solvers presented in
\cite{nukalajpamg1} are superior to iterative solvers in
two-dimensional lattice systems, for 3D lattice systems, the memory
demands brought about by the amount of fill-in during the sparse
Cholesky factorization favor iterative solvers.  The authors have
developed an algorithm based on a block-circulant preconditioned
conjugate gradient (CG) iterative scheme \cite{nukalajpamg2} for
simulating 3D random fuse networks. The block-circulant preconditioner
was shown to be superior compared with the {\it optimal}
point-circulant preconditioner for simulating 3D random fuse networks
\cite{nukalajpamg2}.  Since these block-circulant and {\it optimal}
point-circulant preconditioners achieve favorable clustering of
eigenvalues, these algorithms significantly reduced the
computational time required for solving large lattice systems
in comparison with the Fourier accelerated iterative
schemes used for modeling lattice breakdown \cite{batrouni98,ramstad}.
Using the algorithm presented in
\cite{nukalajpamg2}, we have performed numerical simulations on 3D
cube lattice networks with $L=100$ and $H$ varying from 6 to 100.

\section{Elastic interactions and crack front stability}

Before studying planar crack propagation, it is instructive to
study how the stress is distributed in presence of a crack of
length $a$ in a sample of thickness $H$ considering the two
boundary conditions employed. An analysis of the stress concentration
is useful to assess the stability of the crack under a constant
applied voltage. To this end we define an \textit{enhancement factor} $K \equiv V_{a}/V$,
where $V_{a}$ is the voltage drop across the vertical bonds
ahead of a crack of length $a$ under an applied voltage $V$.
Notice that the enhancement factor $K$ is a discrete version of the stress intensity
factor usually defined in the continuum to quantify the divergence of the stress
ahead of the crack tip. Here, we are working with
a discrete lattice and therefore do not have to worry about singularities.
Another important difference is  that since we are simulating the system imposing a voltage drop
(or displacement), we define $K$ in terms of voltages rather than
currents.

\begin{figure}[th]
\includegraphics[width=1.\linewidth]{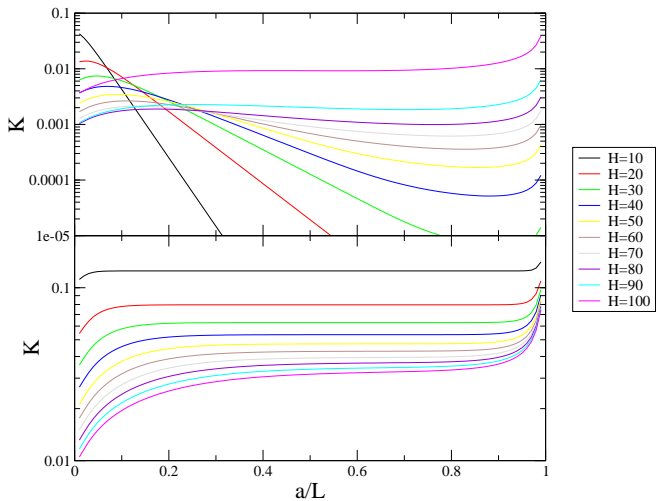}
\caption{The  factor $K$ as a function of the crack length, computed numerically for systems with
different thickness $H$. Here, $K$ is defined as the voltage drop in the bonds ahead of the crack
tip for unit applied voltage. In the top panel, the system is loaded by imposing a constant voltage at
the left edges of the system (line loading). In the bottom panel, the constant voltage is imposed on the
entire top plate (plane loading).}
\label{fig:1}
\end{figure}

\begin{figure}[th]
\vspace{1cm}

\includegraphics[width=.7\linewidth]{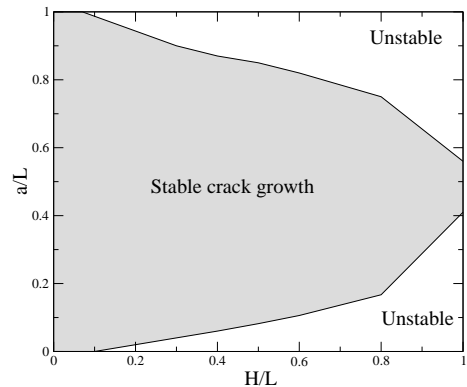}
\caption{The phase diagram of the crack under line loading. The stable region is defined by a
decrease of $K$ as $a$ is increased, implying that the stress in front of the crack
decreases as the crack advances. Under plane loading the crack is always unstable or at most
marginally stable (i.e. $K=const$).}
\label{fig:2}
\end{figure}

To link $K$ to the crack stability, we consider its variation
as a function of the crack length $a$. If $K$ increases with $a$
we expect, on average, an unstable crack growth. This is because as the crack
advances by one step the voltage drop ahead of it increases making
a further failure more likely. This is of course rigorously true for
only for very weak disorder and occasionally one can find a stable crack
even when $K$ increases, due to a particular combination of the random
thresholds. In Fig.~\ref{fig:1} we report $K$ as a function of $a$ for
different values of the relative thickness $H/L$ using the two different
boundary conditions. From this graph one can define the regions of
crack stability by considering the conditions for which $K$ decreases with $a$.
In this way, we see that under plane loading cracks are never stable although
for small $H/L$, we observe a region of marginal stability where $K$ is
roughly constant. On the other hand, under line loading $K$ decreases
exponentially for small $H/L$, leading to stable crack propagation.
Notice however, that for larger $H/L$, at small and large $a$, we
would still expect unstable crack growth. The observations can be
summarized in a phase diagram, Fig.~\ref{fig:2}, where we
report the stable and unstable crack growth regions for line loading.

\begin{figure}[th]
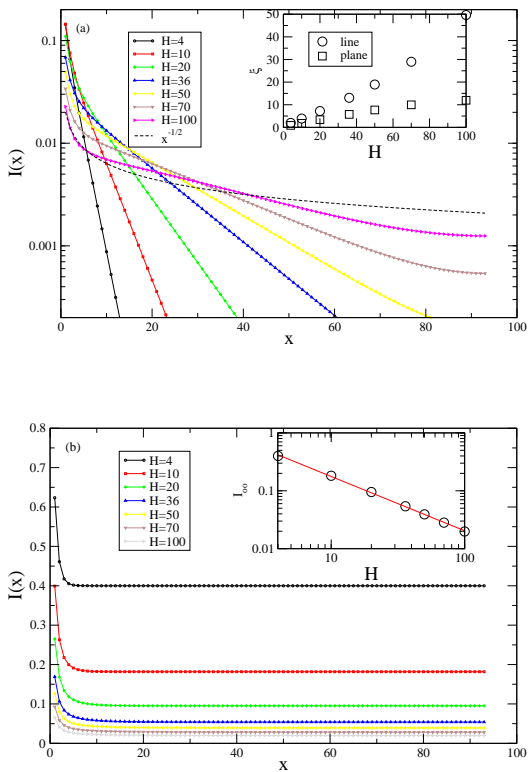

\includegraphics[width=.8\columnwidth]{Itip_line.eps}

\vspace{1cm}

\includegraphics[width=.8\columnwidth]{Itip_plane.eps}

\caption{The vertical current across the weak plane decays exponentially
as a function of the  distance from the crack tip for different thicknesses for both
line and plane loading. (a) For line loading,
the current decays exponentially to zero. A characteristic length $\xi$ can be extracted from
the decay of the currents. As shown in the inset, $\xi$ is a linear function of the thickness $H$, but
it is larger for line loading (i.e. $\xi_{line}\simeq 3\xi_{plane}$) (b) For plane loading,
the current decays exponentially until it reaches a constant value, which decreases as $1/H$ (inset)}
\label{fig:3}
\end{figure}

According to continuum theory in the limit $H\to\infty$, the current ahead of
the crack should decay as $1/\sqrt{r}$. For finite thickness we expect that
a characteristic length emerges \cite{zapperi00}. As shown in Fig~\ref{fig:3}, the current is
found to decay exponentially defining a characteristic length $\xi$. Under line loading, the
current decays to zero, while for plane loading it decays to a value $i_\infty$ that
decreases as $1/H$, vanishing as $H\to \infty$ (see the inset of Fig~\ref{fig:3}b). Deviations
from the exponential behavior can be seen at short distances and large $H$, showing that
the decay is crossing over to the expected $1/\sqrt{r}$ behavior.

\begin{figure}[hbtp]
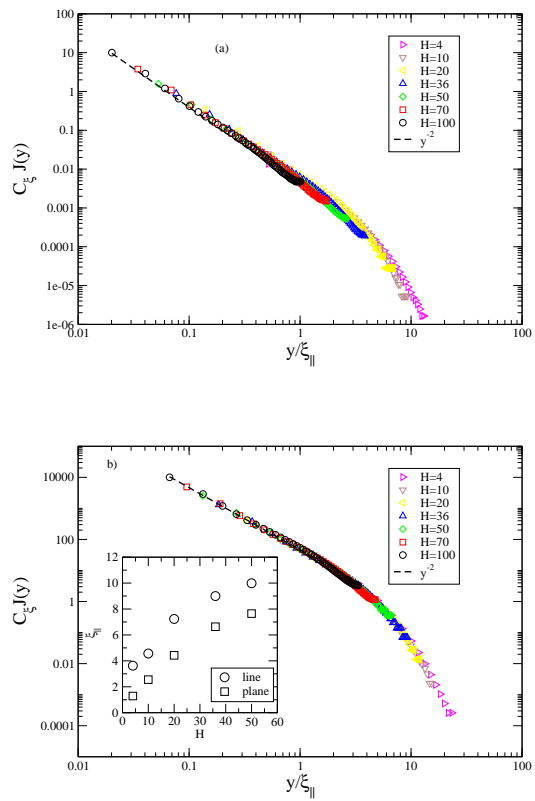

\centerline{\psfig{file=Collapse_kernel_line.eps,width=0.8\columnwidth,clip=!}}

\vspace{1cm}

\centerline{\psfig{file=Collapsed_kernel_plane.eps,width=0.8\columnwidth,clip=!}}
\caption{The variations in the stress intensity factor following the
failure of one bond ahead of the crack as a function of the distance
from the bond parallel to the crack direction. This is equivalent
to a self-interaction kernel, scaling as $1/y^2$ up to a cutoff
length $\xi_\parallel$. Data for different thickness are collapsed
by rescaling the distance by a characteristic length $\xi_\parallel$
whose values are reported in the insets. Results
for line loading (a) and plane loading (b).}
\label{fig:5}
\end{figure}

Since we are interested in planar crack propagation in presence of
disorder, we study the variations in the enhancement factor due to
a small variation in the crack profile. We consider a straight
planar crack of length $a=8$ and remove a single fuse ahead of the
crack. We then compute the increment $J(x)$ of the enhancement
factor $K$ as a function of the distance from the removed fuse.
This function is closely related to the first-order variation of the stress intensity
factors, computed by Gao and Rice \cite{gao89} and commonly employed in
line models for planar crack front propagation \cite{schmittbuhl95,ramanathan97,ramanathan98}.
Based on this analogy, we can expect that in the limit $H\to\infty$ it should
be $J(x) \propto x^{-2}$ \cite{gao89}. This results is confirmed
by our simulations, reported in Fig.~\ref{fig:5}, showing that the
a finite thickness $H$ induces again a characteristic length $\xi_{||}$. The data
obtained for different $H$ can be collapsed according to the scaling
form $J(x) = x^{-2} f(x/\xi_{||})$, where $f(x)$ decays exponentially and $\xi_{||}$ is obtained from a fit.
In the case of plane loading, we find that $\xi_{||}$ depends on
$H$ and goes linearly to zero as $H \to 0$. For line loading, however, the
characteristic length $\xi_{||}$ depends also on the crack size $a$ and therefore
it does not go to zero as $H\to 0$ (see Fig \ref{fig:6})

\vspace{1cm}

\vspace{1cm}

\section{Crack front roughness and avalanches}

\begin{figure}[htbp]
\centerline{\psfig{file=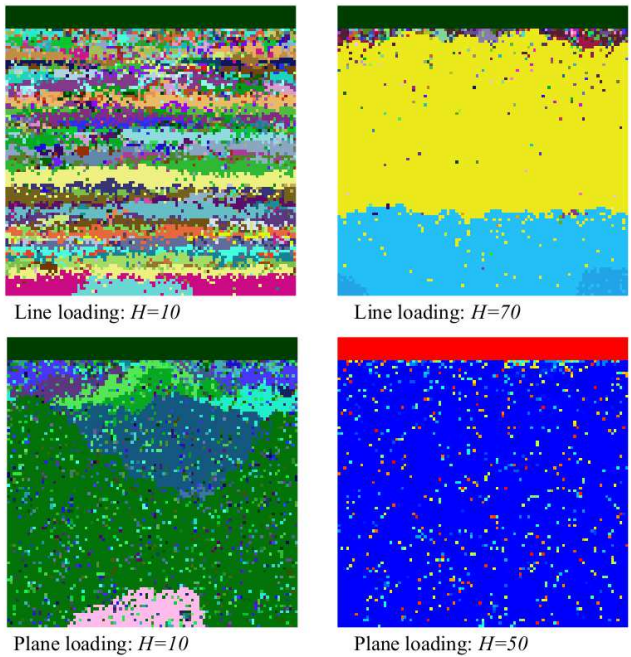,width=\columnwidth,clip=!}}
\caption{The avalanche progression as a function of the loading model and the sample thickness.
Different avalanches are identified by random colors.}
\label{fig:6}
\end{figure}

As we load the system, the crack advances but due to the presence
of disorder in the breaking thresholds the crack front roughens and
dynamics is composed by a sequence of avalanches (Fig.~\ref{fig:6}),
in close analogy to what is observed in experiments \cite{schmittbuhl97,delaplace99,maloy01,maloy03,maloy06,santucci10}.

\begin{figure}[tbph]
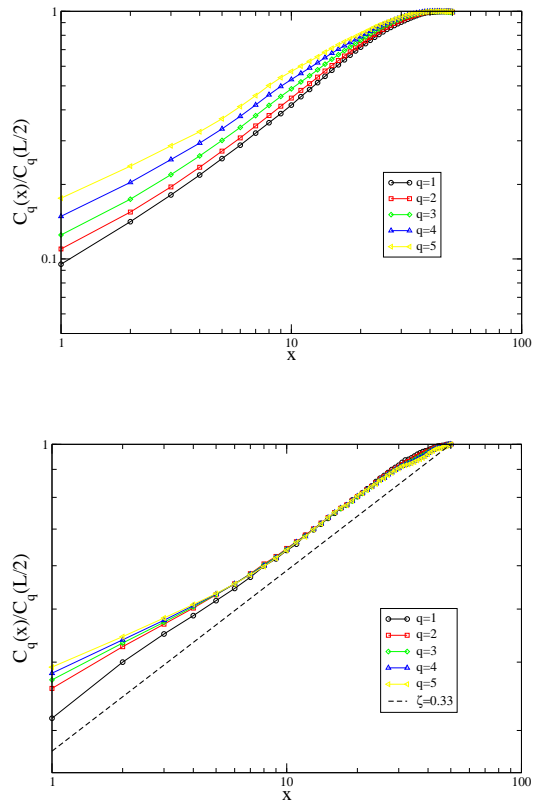

\centerline{\psfig{file=Cq6.eps,width=0.8\columnwidth,clip=!}}
\vspace{1cm}
\centerline{\psfig{file=Cq50.eps,width=0.8\columnwidth,clip=!}}
\caption{The $q-$ moments of the correlation functions for cracks under plane loading for H=6 (top) and  H=100 (bottom). The dashed line indicates a power law with exponent $\zeta=0.33$.}
\label{fig:7}
\end{figure}

To quantify the fluctuations in the crack morphology as a function of the sample thickness,
we follow the  multiscaling analysis commonly employed to study fracture
fronts \cite{bouchbinder06,alava06b,santucci07,santucci10}
and compute the $q-$moments of the correlation function
\begin{equation}
C_q(x) = (\langle h(x'+x)h(x') \rangle)^{1/q},
\end{equation}
where $h(x)$ is the position of the front. We perform the average over different
realizations of disorder and consider only cracks located in the central part of the
lattice, to avoid boundary effects. The results are illustrated in Fig \ref{fig:7}
where we show that for large thickness (i.e. a cubic system with $H=L$)
all the moments scales as $x^\zeta$, with $\zeta \simeq 0.33$ on large length scales
and an indication of multiscaling behavior at small lengthscales (Fig \ref{fig:7}a).
This is very similar to what is  found in experiments \cite{santucci10}. At low
thicknesses, however, we observe a wide multiscaling regime over all the
available lengthscales (Fig \ref{fig:7}b). This result indicates that the sample
thickness controls the crossover scale between the short-scale multiscaling
regime and the large-scale interface depinning scaling.

The nature and morphology of the avalanches depends on the loading mode
and the thickness as shown in Fig.~\ref{fig:6}. Under line loading and
for a small thickness, crack line motion is hindered by a strong restoring
force which limits the avalanche size. For a large thickness and for
plane loading, the front dynamics is unstable and therefore we observe
large avalanches that span a considerable fraction of the system together
with other smaller avalanches.

\begin{figure}[hbtp]
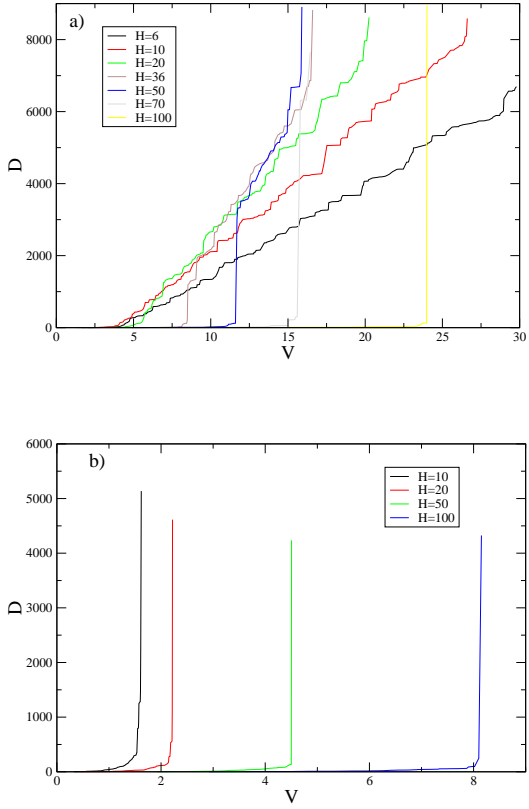

\centerline{\psfig{file=damage2.eps,width=0.8\columnwidth,clip=!}}
\vspace{1cm}
\centerline{\psfig{file=damage.eps,width=0.8\columnwidth,clip=!}}

\caption{The accumulated damage as a function of the applied voltage for a) line loading
and b) plane loading. The staircase character of the curve is a signature of avalanche
behavior.}
\label{fig:8}
\end{figure}

\begin{figure}[hbtp]
\centerline{\psfig{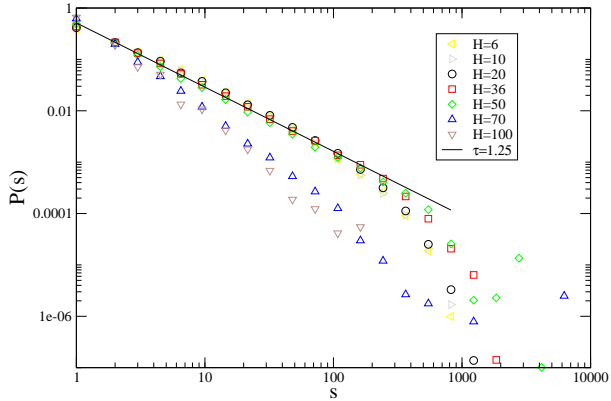}}
\caption{The avalanche size distribution measured under line loading for different values
of the sample thickness.}
\label{fig:9}
\end{figure}

The progression of the avalanches can be observed in Fig \ref{fig:8}
where we report the total lattice damage $D$, defined as the number
of broken bonds, for typical realizations of the simulations. $D$ illustrates nicely the effect of the stability analysis from above on avalanches. Under line loading and low thickness, we observe a sequence of random avalanches with wide size distribution. For larger thickness, we observed the nucleation of large avalanches which correspond to unstable crack growth  (see  Fig \ref{fig:8}a). The role of instability
is even more apparent under plane loading we see large system spanning avalanches (see  Fig \ref{fig:8}b).
The distribution of avalanche sizes for line loading is reported in Fig. \ref{fig:9}. For small 
and intermediate thickness we observe a power law distribution with exponent $\tau \simeq 1.25$
and a cutoff that increases with the thickness. The measured exponent is in agreement with the
result expected for the crack line depinning model \cite{bonamy08,laurson10}. For larger thickness,
we see a deviation from this result and the power law exponent becomes much larger ($\tau \simeq 2$)
and the distribution displays a peak at large avalanches which is a signature again of the large
unstable avalanches.

\section{Strength distribution and size effects}

\begin{figure}[hbtp]
\centerline{\psfig{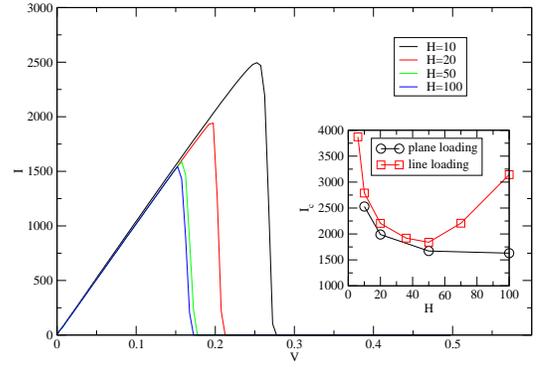}}
\caption{The voltage-current curve for the model under plane loading. The inset shows the fracture current
as a function of the thickness $H$ for plane and line loading.}
\label{fig:10}
\end{figure}

In Fig. \ref{fig:10} we report the voltage-current curves obtained in the model under planar loading for different thickness values. These curves are related to the stress strain curves by defining shear stress
as $\sigma\equiv I/L^2$ and shear strain as $\epsilon \equiv V/H$. The inset of Fig. \ref{fig:10} displays
the size effect for plane and line loading. While for plane loading the strength decreases with the thickness,
for line loading the strength increases at large thickness. This crossover is due to the fact that the planar
crack becomes unstable at larger $H$ as also illustrated in Fig. \ref{fig:8}.

\begin{figure}[hbtp]
\centerline{\psfig{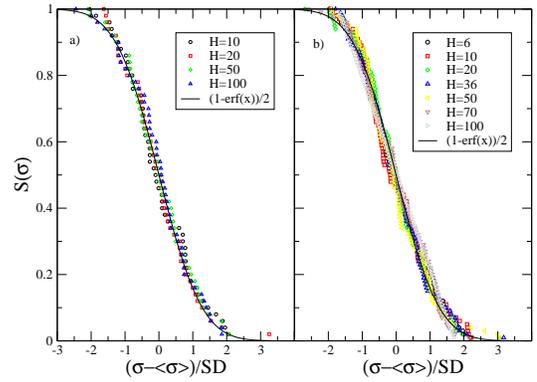}}
\caption{The stress survival distribution for a) plane and b) line loading as a function for
the sample thickness. Data are plotted as a function of reduced variables 
$(\sigma-\langle \sigma\rangle)/SD$ and the corresponding Gaussian distribution
is reported for comparison.}
\label{fig:11}
\end{figure}

We also measure the stress survival distribution $S(\sigma)$ defined as the probability that the
sample does not fracture at stress $\sigma$. In both cases the distribution is well described by 
Gaussian statistics as it is shown in Fig. \ref{fig:11} by using reduced variables $(\sigma-\langle \sigma\rangle)/SD$, where $\langle\sigma\rangle$ and $SD$ are the average and standard deviation
of $\sigma$. The presence of Gaussian statistics is expected in systems that have one dominating
crack so that statistical size effects described by extreme value theory are not present \cite{alava09}.

\section{Conclusions}
Planar crack propagation has been for some time a test ground for theories of depinning in the context of fracture, and much progress has been made. For understanding the connections between the paradigm of a non-equilibrium critical point for a driven crack and actual behavior in an experiment, it is necessary to investigate in a general manner the effects of loading conditions and sample geometry. Usual theory accounts for the distance the crack propagates, and for the finite length of the crack line and provides predictions. Here, we have added the effect of the finite sample thickness, which influences among others the effective form of the interactions along the crack ("elastic kernel"). 

The loading has also been found to be of importance, and the comparison between the line and plane loading cases, where the former is close to most recent experiments, in fact shows major differences. The coarse-grained stability properties of the "experiment" are decisive for the presence of the collective phenomena, ie. avalanches. We would think that this hints of a need for further investigations of other possible loading protocols. Generally, we find also the signatures of the universality class of long-range elastic line depinning: avalanches with the expected size distribution, and line roughening with a roughness exponent as expected. These observations should have also an impact on understanding the design of interfacial layers for adhesive properties or for fracture toughness.

\section*{Acknowledgments}
SZ is supported by the European Research Council Advanced Grant 2011 - SIZEFFECTS and thanks the visiting professor program of Aalto University, School of Science and the Aalto Science Institute.
MJA thanks the support from the Academy of Finland through the COMP Center of
Excellence. Part of this work was performed when PKVVN was a senior researcher at Oak Ridge National Laboratory. PKVVN acknowledges the support received from the Mathematical, Information and Computational Sciences Division, Office of Advanced Scientific Computing Research, U.S. Department of Energy under Contract No. DE-AC05-00OR22725 with UTBattelle,LLC.

\end{document}